\documentclass[a4paper]{PoS}

\title{Searching Sub-Millisecond Pulsars in Accreting Neutron Stars}

\ShortTitle{Sub-ms pulsars}

\author{\speaker{Alessandro Patruno}\\
        Astronomical Institute ``A. Pannekoek'', University of Amsterdam, 1098XH Science Park 904, Amsterdam, The Netherlands\\
        E-mail: \email{a.patruno@uva.nl}}

\abstract{Measuring the spin of Accreting Neutron Stars is important
because it can provide constraints on the Equation of State of
ultra-dense matter. Particularly crucial to our physical understanding
is the discovery of sub-millisecond pulsars, because this will
immediately rule out many proposed models for the ground state of
dense matter.  So far, it has been impossible to accomplish this
because, for still unknown reasons, only a small amount of Accreting
Neutron Stars exhibit coherent pulsations.  An intriguing explanation
for the lack of pulsations is that they form only on neutron stars
accreting with a very low average mass accretion rate. I have searched
pulsations in the faintest persistent X-ray source known to date and
I found no evidence for pulsations. The implications for accretion theory
are very stringent, clearly showing that our understanding of the
pulse formation process is not complete. I discuss which sources are
optimal to continue the search of sub-ms pulsars and which are the new
constraints that theoretical models need to explain to provide a
complete description of these systems. }

\FullConference{High Time Resolution Astrophysics IV - The Era of Extremely Large Telescopes - HTRA-IV,\\
		May 5-7, 2010\\
		Agios Nikolaos, Crete, Greece}

\begin{document}

\section{Introduction}

Neutron stars are objects of great astrophysical importance because
the density in their cores can reach values several times higher than
normal nuclear matter. At such ultra-high densities the behaviour of
matter is not well known, and exotic nuclear processes may take place
along with the formation of new states of matter~\cite{weber, glen97}.
The properties of ultra-dense matter are determined by the equation of
state (EoS) of neutron star matter~\cite{lattimer}. This EoS is
relatively well known at values below the saturation density of normal
nuclear matter, but is extremely uncertain above this threshold. The
possibility of constraining the EoS of ultra-dense matter can be
achieved only by studying neutron stars and is fundamental for our
comprehension of nuclear and sub-nuclear interactions.

\begin{figure}[t]
  \begin{center}
    \rotatebox{0}{\includegraphics[width=0.6\textwidth]{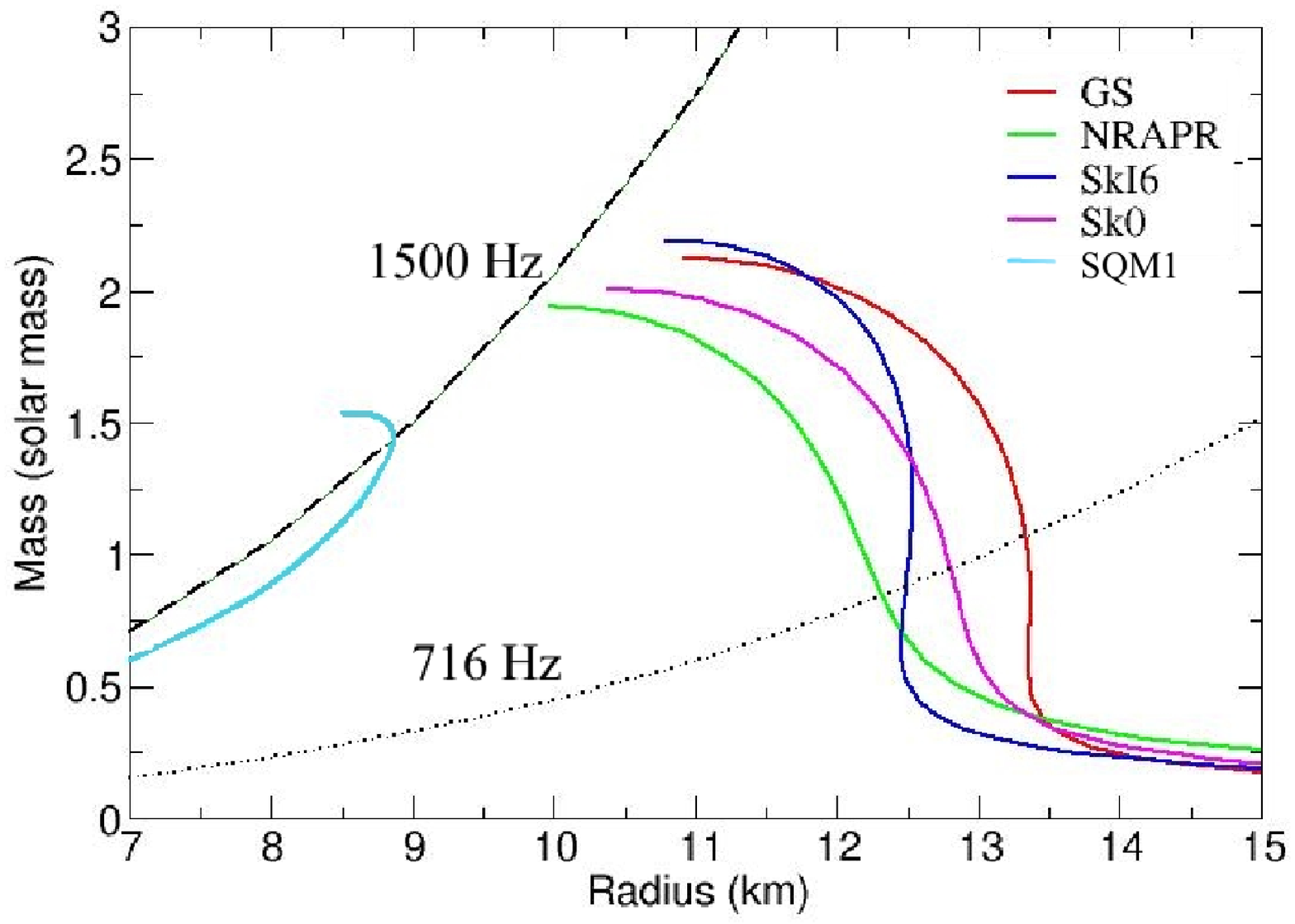}}
  \end{center}
  \caption{Mass-Radius relation for neutron stars with different 
Equation of State (see~\cite{lattimer} for a description 
of the EoS used). The dotted and dashed lines show the constraints that 
a neutron star spinning at 1500 Hz and 716 Hz would imply on the EoS: 
any point below these lines cannot represent the Mass and Radius of the
spinning neutron star, since the centrifugal forces would imply mass
shedding and disruption of the object.}
\label{fig1}
\end{figure}

Neutron stars exist as a wide variety of objects that can be studied
in different electromagnetic wavelengths, ranging from radio to gamma
rays. A particularly interesting way to constrain the EoS is via the
search of neutron stars spinning at sub millisecond periods.  The
reason for this is that neutron stars cannot have arbitrarily short
spin periods, because the centrifugal force would overcome the
effective gravity and induce mass shedding. The minimum achievable
spin period of a neutron star depends on its particular EoS because
repulsive nuclear interactions have a fundamental role in partially
balancing the gravitational pull. It is also possible that self-bound
stars exist, like strange stars, who are kept together by the strong
interaction and whose rotational period can reach a much smaller value
than hadronic neutron stars~\cite{glen97}.  Therefore by determining
the minimum rotational period it is possible to exclude a large amount
of EoS and understand for example the degree of stiffness of
ultra-dense matter (see Fig.~\ref{fig1}).  To date, the fastest known
neutron star is the radio pulsar PSR J1748-2446ad, spinning at 716 Hz
($\sim1.4$ ms~\cite{hessels}).  Although this period corresponds to a
rotational velocity of $\sim15\%$ the speed of light (assuming a
neutron star radius of 10 km), it is still compatible with basically
all realistic EoS. Radio pulsars with spin periods beyond the 1.4 ms
limit are difficult to detect due to observational limitations like
the unknown dispersion measure. In X-rays instead the search for
pulsation is not biased at least down to 0.5 ms and in this sense the
subclass of accreting neutron star is the optimal target for sub-ms
pulsar searches.
 
The most interesting accreting neutron stars are those in which the
companion star has a low mass: they are called neutron star low mass
X-ray binaries (NS-LMXBs). These systems are very old (with ages of the
order of $10^{8}-10^{9}$ yr) and the neutron star has spent a
substantial fraction of its life accreting gas and spinning up via
transfer of angular momentum~\cite{tauris}.  This means that there is 
high chance of finding a millisecond pulsar, and possibly a sub-ms
pulsar, in a NS-LMXB. In this work I present the
first results of a deep pulse search in these kind of X-ray binaries.

\section{Accreting Neutron Stars}

The number of known NS-LMXBs in our galaxy is $\sim100$ (see for
example~\cite{liu}).  There are also many binaries in which the
companion is a high mass star, but they are of little interest here
because their relatively young age and the intense magnetic field of
the neutron star prevent them to having been spun up to the
millisecond range ~\cite{bildsten}.

In NS-LMXBs instead, the majority of neutron stars is old, and their
magnetic field is believed to have decayed to values around
$10^{8}-10^{9}$ G.  Some gas leaves the companion via Roche Lobe
Overflow and forms an accretion disc around the neutron star.  When
the gas loses enough angular momentum and arrives at the innermost
regions of the disc, two phenomena can take place:
\begin{itemize}
\item if the neutron star
has an external magnetic field larger than $\sim10^{7}-10^{8}$ G, the 
gas is (partially) channeled
towards the magnetic poles 
\item if the neutron star has no relevant external magnetic field, then 
the gas keeps spiraling down to the innermost stable circular orbit (ISCO) 
and then plunges on the neutron star surface. If the ISCO is below the 
neutron star surface, as it is the case for some EoS, then the disc
extends down to the neutron star surface. 
\end{itemize}

In the case of channeled accretion, the gas might flow in the magnetosphere
in ordered structures producing two sites of intense X-ray radiation
at the magnetic poles. If the magnetic and rotational axes are
misaligned, the X-rays are modulated at the spin period of the neutron
star and are called ``accreting X-ray pulsars''. In the second scenario the
accreting gas does not produce any sufficient emission asymmetry and no
pulsations are produced.

\section{Accreting Millisecond X-ray Pulsars}

Among the $\sim100$ NS-LMXBs, only a small number show X-ray
pulsations.  A very small sub-group is composed by NS-LMXBs with very
long spin periods (tens to hundreds of seconds, see~\cite{bildsten}),
while 13 systems show X-ray pulsations in the millisecond range
(Accreting Millisecond X-ray Pulsars, AMXPs, see Table 1).

All Accreting Millisecond X-ray pulsars share similar characteristics: 
\begin{itemize}
\item pulsations are sinusoidal, with little or no harmonic content
\item the companion star has a low mass from $1M_{\odot}$ down to 
0.007$M_{\odot}$
\item the binary orbital period is short, between 0.7 and 20 hr.
\end{itemize}

The rotation of the neutron star modulates the X-ray emission and
gives a sinusoidal shape for the pulses, mainly because of a projected
area effect.  The fact that all AMXPs have a low mass companion star
is a consequence of binary evolution: if the companion star were
massive, then accretion would take place either via an intense wind or
via a dynamical/thermal timescale that would rapidly reduce the mass of the
companion and possibly engulf the whole system. The mass transfer becomes
stable when the mass ratio between the donor and the neutron star
becomes close or smaller than $\sim1$. Once a stable mass transfer
sets in, the neutron star is spun up on timescales of $10^{7}-10^{8}$
yr, which are rather short when compared to the lifetime of the binary
($\sim10^{9}$ yr, see~\cite{tauris}).

To have a sufficiently intense mass transfer from a low mass companion
it is necessary to have Roche Lobe Overflow, because the stellar wind
is too weak to provide enough gas for the production of intense X-ray
pulses. Since the companion stars have usually a mass much smaller than 
 $1\,M_{\odot}$, reaching the impressively small value of $\sim7$ Jupiter
masses (see Table 1), they are usually compact or 
ultra-compact systems: to overflow the Roche Lobe it is necessary that
the companion is very close to the neutron star~\cite{tauris}.

\begin{table}
\caption{Accreting Millisecond X-ray Pulsars}

\scriptsize
\begin{center}
\begin{tabular}{lccccccc}
\hline
\hline
Source & $\nu_{s}$ & $P_{orb}$ & $f_{x}$  & $M_{c,min}$  & Type I Bursts & B-O & Reference\\
 & (Hz) & (min) & ($M_{\odot}$) & ($M_{\odot}$) &  & & \\
\hline
SAX J1808.4--3658\dotfill & 401 & 121 & $3.8\times 10^{-5}$ & 0.043 & Yes & Yes & \cite{hartman}\\
XTE J1751--305\dotfill & 435 & 42.4 & $1.3\times 10^{-6}$ & 0.014 & No & No & \cite{papitto08}\\
XTE J0929--314\dotfill  & 185 & 43.6 & $2.9\times 10^{-7}$ & 0.0083 & No & No & \cite{galloway02}\\
XTE J807--294\dotfill  & 190 & 40.1 & $1.5\times 10^{-7}$ & 0.0066 & No & No & \cite{patruno10A} \\
XTE J1814--338\dotfill  & 314 & 257 & $2.0\times 10^{-3}$ & 0.17 & Yes & Yes & \cite{papitto07}\\
IGR J00291+5934\dotfill  & 599 & 147 & $2.8\times 10^{-5}$ & 0.039 & No & No & \cite{patruno10}\\
HETE J1900.1--2455\dotfill  & 377 &  83.3 & $2.0\times 10^{-6}$ & 0.016 & Yes & Yes & \cite{kaaret06}\\
Swift J1756.9--2508\dotfill  & 182 & 54.7 &  $1.6\times 10^{-7}$ & 0.007 & No & No & \cite{patruno10B}\\
Aql X--1\dotfill     & 550 & 1137 & N/A & N/A & Yes & Yes & \cite{casella}\\
SAX J1748.9--2021\dotfill  & 442 & 522 &  $4.8\times 10^{-4}$ & 0.1 & Yes & No & \cite{patruno09}\\
NGC6440 X-2\dotfill & 206 & 57 & $1.6\times 10^{-7}$ & 0.0067 & No & No & \cite{altamirano10A}\\
IGR J17511-3057\dotfill & 245 & 208 & $1.1\times 10^{-3}$ & 0.13 & Yes & Yes & \cite{papitto10}\\
Swift J1749.4-2807\dotfill & 518 & 529 & $5.5\times 10^{-2}$ & 0.59 & No & No & \cite{altamirano10B}\\
\hline
\end{tabular}\\
$\nu_{s}$ is the spin frequency, $P_{orb}$ the orbital period, $f_{x}$
is the X-ray mass function, $M_{c,min}$ is the minimum companion mass.
The 6th and 7th columns identify bursting sources and those that show
Burst-Oscillations (B-O). The last column shows references for the
timing parameters. The references chosen are those in which the most
precise timing parameters are reported.
\end{center}
\label{tab:AMXP}
\end{table}

\section{Why Does Only a Small Number of Accreting Neutron Stars Pulsate ?}

A consequence of standard
accretion theory and of the \textit{recycling scenario}~\cite{alpar}
is that all NS-LMXBs that are accreting at a sufficiently high rate
must show X-ray pulsations.  However, as anticipated in the previous
sections, only 13 out of $\sim100$ NS-LMXBs do pulsate.  The reason
for this is still unknown.  Many models exist today that try to
provide an answer to this problem, but none has been considered
conclusive yet. Some models suggest that the dipolar magnetic field is
buried in the neutron star crust~\cite{cumming}, thus preventing any
channeled accretion. Other models predict a smearing of pulsations due
to gravitational lensing~\cite{wood}, or to scattering in a hot corona
surrounding the pulsar~\cite{titarchuk} or to alignment between the
spin and the magnetic axes~\cite{lamb}. Other models have proposed the
onset of magneto-hydrodynamical instabilities, that would develop at
the magnetosphere-disc boundary, like Rayleigh-Taylor instabilities
that destroy the coherence of the accretion~\cite{kulkarni}.  It is
beyond the scope of this contribution to explain such models, but what
is important to stress once more is that none of them has found a final
confirmation yet.

A major breakthrough in our comprehension of the pulse formation
process was achieved with the discovery of intermittency in some
AMXPs. The first reported episode of intermittency was done
for the AMXP HETE J1900.1-2455~\cite{kaaret06, galloway07}. This pulsar
showed approximately 2 months of continuous pulsations during an
outburst that started in 2005. The pulsations then disappeared and
were not detected for the following 5 years, up to the time of writing this
contribution. Two other intermittent sources were then discovered (Aql
X-1 and SAX J1748.9-2021, see Table 1). The most remarkable was Aql
X-1: a single 120 s long episode of pulsations was detected with high
significance among more than 1.2 Ms of observations. SAX
J1748.9-2021 showed instead repeated episodes of intermittency with
pulsations appearing and disappearing on timescales of a few hundred
seconds and with a possible connection with the occurrence of Type I
X-ray bursts (see Table 1 and references therein). These three sources
are of great importance because they might bridge the gap between
AMXPs and non-pulsating NS-LMXBs. If the mechanism that produces
intermittency is also responsible for the lack of pulsations in many
NS-LMXBs, then any pulse formation model must be able to explain the
existence of rapid timescales as short as $\sim100$ s for the formation
and destruction of X-ray pulses.

\section{Nuclear Powered X-ray Pulsars}

The gas accreted on the neutron star surface accumulates and might
give rise to thermonuclear explosions if specific temperature and
density conditions are met. Thermonuclear explosions (a.k.a. Type I
X-ray bursts) are sudden releases of nuclear energy that happen on
timescales of a few tens to hundreds of seconds.  Although the energy
released per accreted baryon ($\sim200$MeV/baryon) is much larger than
the nuclear energy released during stable burning ($\sim5$MeV/baryon),
the unstable nuclear reactions taking place during Type I X-ray burst
release energy that completely dominates the production of X-ray
radiation. During Type I X-ray bursts, nuclear powered oscillations at
(or very close to) the spin frequency of the neutron star are sometimes
observed~\cite{strohmayer}.
  
The precise mechanism for the formation of burst oscillations is still
unknown and current models propose a relation with global modes of
oscillation of the neutron star surface, thermonuclear hurricanes or
hot-spot emission asymmetries due to the thermonuclear flame spreading
(see for example~\cite{watts} and references therein).

We know 11 NS-LMXBs that show burst oscillations, and they are of high
importance because 6 of these have never shown accretion powered
pulses; therefore their spin would have never been known.  
Considering nuclear powered and accretion powered
oscillations, we have a sample of 23 NS-LMXBs with known confirmed
spin periods,which means that now we begin having a good sample for
accreting neutron star spin distribution.  In a recent
paper~\cite{patruno10}, I reported the updated spin frequency
distribution of the 23 known accretion and nuclear powered millisecond
X-ray pulsars. The distribution shows still a sharp cutoff around 730
Hz with 99\% confidence level and with median value for the spin
frequency of 415 Hz. The spin distribution cutoff limit of 730 Hz
still resists after about 7 years and with a sample size doubled from
its first calculation~\cite{chakrabarty}.  The only claimed sub-ms
pulsar is the 1122 Hz accreting neutron star XTE
J1739-285~\cite{kaaret}. However, this result is still unconfirmed.

Since all persistent AMXPs have shown a rather low average mass
accretion rate, it is well possible that the key propriety that
determines why these sources show pulsations while other NS-LMXBs
do not is indeed the low average mass accretion rate.  It has been
proposed that the dipolar magnetic field of neutron stars is buried in
the crust during accretion~\cite{cumming}.  
If the Ohmic diffusion timescale, that regulates the diffusion of the
magnetic field in the plasma, is shorter than the accretion timescale,
then the magnetic filed is able to spread outside the neutron star
surface and affect the accretion process in the disc.  The buried
magnetic field model is one of the possibilities to explain the lack
of pulsations in the majority of NS-LMXBs.  However, a considerable
theoretical effort needs to be still made, because the current model
rely on several approximations that might considerably affect the
results. A simple prediction of this model, which seems to be at the
origin of its success, is that the magnetic field is buried when the
mass accretion rate is typically above $2\%$ of the Eddington limit.

It seems that all AMXPs have a low long-term average mass accretion
rate~\cite{heinke, ozel}, which is much smaller than that of non pulsating NS-LMXBs,
suggesting that this is the right direction to go when searching for
pulsators among NS-LMXBs. However, some care has to be taken when
calculating the long term average mass accretion rate of AMXPs, since
the distance and the recurrence time of many of them is rather
uncertain.

To verify the consistency of the buried magnetic field model it is
therefore crucial to search for pulsations in those NS-LMXBs that have
a very low average mass accretion rate.  Recently, a new sub-class of
LMXBs has been discovered, called Very Faint X-ray Sources (VFXSs)
whose main property is to have very faint peak luminosities always
below $10^{36}\rm\,erg\,s^{-1}$~\cite{wijnands}.  If the accretor is a
neutron star, then these are the optimal targets for any pulse search
campaign.

\section{The Very Faint X-ray Source 1RXS J171824.2-402934}

The faintest persistent VFXS known to date is 1RXSJ171824.2-402934.
The source shows Type I X-ray bursts with photospheric radius
expansion. Assuming a nearly Eddington limited emission during the
luminosity peak, a distance of $6.5\pm 0.5$kpc was
inferred~\cite{kaptein}.  A {\it Chandra} observation carried in
2004~\cite{in'tzand05} has shown a persistent emission with a total
unabsorbed $0.5-10$keV flux of $(9.7\pm1.7)\times
10^{-12}\rm\,erg\,cm^{-2}\,s^{-1}$ which corresponds to a luminosity
of $4.9\times 10^{34}\rm\,erg\,s^{-1}$, for a distance of 6.5 kpc.
The X-ray luminosity is approaching $10^{-4}L_{Edd}$ and the inferred
mean mass transfer rate is $\sim9\times
10^{-12}\rm\,M_{\odot}\,yr^{-1}$, both are in perfect agreement with the
requirement set by the buried magnetic field model to produce
pulsations. If the neutron star atmosphere is strongly contaminated by
helium material, the distance obtained from the Type I burst should be
increased up to 9 kpc. Even in this case the luminosity is very faint
($9.4\times 10^{34}\rm\,erg\,s^{-1}$) and the considerations remain
unchanged.  
The spectrum of the source has been fitted with an
absorbed power law with a photon index $\Gamma=2.09^{+0.22}_{-0.24}$
and a column density $N_{H}=1.32^{+0.16}_{-0.12}\times
10^{22}\rm\,cm^{-2}$ which seems to be typical of NS-LMXBs.

In a recent work~\cite{in'tzand09}, it was proposed that this source
has the unique characteristic of being the shortest ultra-compact
binary known to date: its orbital period should be less than 7 min,
compared with the 11 min of the X-ray binary 4U 1820-30.  The
determination of the period is very uncertain, and has to be taken
with caution, but it is interesting anyway to see that the discovery
of pulsations would allow not only a test of magnetic field evolution
theories and the determination of the neutron star spin but also the
confirmation of the shortest orbital period known.
\begin{figure}[t]
  \begin{center}
    \rotatebox{0}{\includegraphics[width=1.0\textwidth]{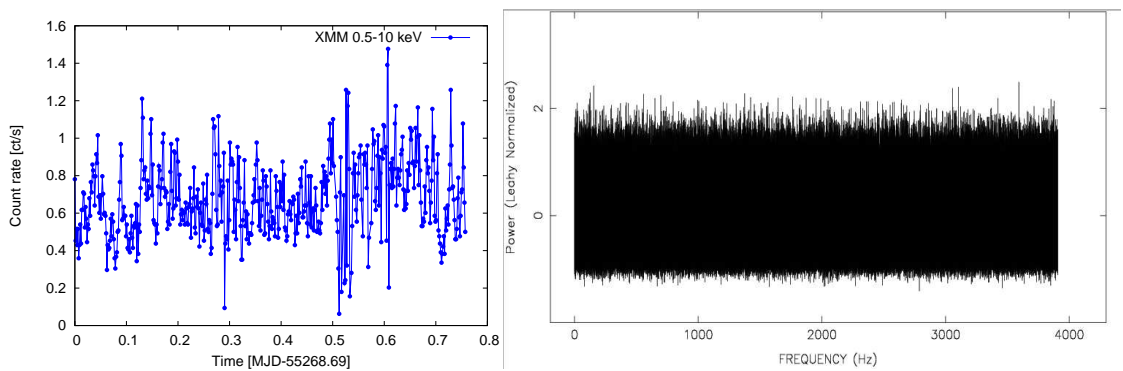}}
  \end{center}
  \caption{X-ray count-rate (left panel) and power spectrum (right panel)
of the Very Faint Source 1RXSJ171824.2-402934. The data refer to an XMM-Newton
observation of 65 ks long data in the 0.5-10 keV energy band. 
No significant X-ray flux variations are observed, and a simple power spectrum
shows no relevant features.}
\end{figure}
I obtained a 65 ks long data from a pointed XMM-Newton
observation of this source, taken with the EPIC-pn camera set up in timing
mode to allow a search for pulsations. The observed source flux is
indeed very low, and never goes above 1.5 ct/s in the 0.5-10 keV
energy band.  This can be translated in an observed average mass
accretion rate of about $6\times\,10^{-12}\rm\,M_{\odot}\,yr^{-1}$
(which is $0.02\%$ Eddington) for an assumed distance of 6.5 kpc and
using spectral parameters as reported in~\cite{in'tzand05}.  The
lightcurve and the power spectrum of the XMM observation are reported
in Figure 2.

An inspection of the power spectrum of the lightcurve shows no evident
pulsations at frequencies below 2 kHz.  Given the background and
source count rates, if a signal with an rms amplitude larger than 2\%
were present, then a power spectrum would show a spike with a
signal-to-noise ratio of at least 5. However, this search technique
makes the non-trivial assumption that the signal stays in one single
Fourier frequency bin during the whole observation.  If the suggestion
of a short period made in~\cite{in'tzand09} is correct, then the
power spectrum is heavily affected by Doppler shifts due to the
orbital motion of the neutron star around the companion and that would
spread the signal in a large number of Fourier frequency bins.

To take this possibility into account I performed a deep pulse search
by using the sideband-search technique explained by~\cite{jouteux}
and~\cite{ransom}.  This search technique is optimal when the length
of the time series is much larger than the orbital period, because a
family of sidebands in the Fourier power spectrum will appear around
the pulsar spin frequency.  An important requirement to increase the
sensitivity is that many orbital periods are included in the
observation so to create a large number of sidebands. The result of
the search did not return any significant spin frequency
candidate. Although the calculation of proper upper limits is strongly
dependent on the assumed orbital and spin parameters during the
search, preliminary limits of less than 1\% rms on the pulse amplitude
can be placed (see~\cite{vaughan}).  Acceleration searches and
calculations of more stringent upper limits are currently under way.

Upper limits of 1\% rms are already very constraining for all theories
of pulse formation, and they question the validity of the argument
that a low value of average mass accretion rate favours the appearance
of pulsations. It remains to be investigated whether the persistence
of 1RXSJ171824.2-402934, and therefore the continuous accretion of
mass, might have a role on the pulse formation process. Therefore it
would be interesting to search for pulsations also in VFXSs
which are transients instead of persistent. 

\section{Theoretical implications}

The lack of pulsations in the VFXS 1RXSJ171824.2-402934 has several
theoretical implications.  First, it shows that a pulse search program
to unveil the existence of sub-ms pulsars in NS-LMXBs is more
difficult than predicted. The existence of sub-ms pulsars has not been
demonstrated so far and deep X-ray data searches have not returned
positive results (see for example~\cite{dib} for a deep search in the
NS-LMXB 4U 1820-30). In a recent paper~\cite{patruno10}, I discussed
how the only two AMXPs for which a long term spin period evolution is
constrained (SAX J1808.4-3658 and IGR J00291+5934) do not show any
evidence for a short-timescale spin evolution. The AMXP SAX
J1808.4-3658 shows a long term \textit{spin-down} over an observed
baseline of 10 years, while IGR J00291+5934 shows the spin increasing
at a rate that would bring it in the sub-ms range in $\sim10$ Gyr.  If
this is a common behaviour among NS-LMXBs, it might imply that none of
the NS-LMXBs have been spun up for a sufficiently long time to reach
sub-ms periods. The lack of sub-ms pulsars might therefore be a simple
consequence of binary evolution (see~\cite{deloye} and references
therein). To establish whether sub-ms pulsars exist or not, it might
be interesting therefore to determine the long term evolution of other
AMXPs, although this might be prohibitive given the long
recurrence time between outbursts of several of them.  A possible
strategy to overcome this problem would be to increase the
sample of known AMXPs via a deep search of targeted candidates, like
the VFXS, to determine whether the effect of a low mass accretion rate
is indeed an important parameter for the pulse formation.

A second consequence of the study of 1RXSJ171824.2-402934 is that a
low average mass accretion rate is not sufficient to guarantee the
formation of X-ray pulses by itself. Recently, the buried magnetic
field model was further developed~\cite{cumming08} with encouraging
results: the external dipole magnetic field can be reduced on
timescales of months and the amount by which the magnetic field in the
neutron star ocean changes depends on the average mass accretion rate
\textit{and also} on the total amount of mass accreted.  This would in
principle explain why 1RXSJ171824.2-402934 does not show pulsations:
if the amount of mass accreted is sufficiently large, then pulsations
might not form.

\acknowledgments{I would like to thank 
R. Wijnands and N. Degenaar for several stimulating discussions
on the Very Faint Source properties, A. L. Watts. and Y. Cavecchi
for discussions on the Equation of State of ultra-dense matter
and thermonuclear burst models.}

\end{document}